\documentclass[lettersize,journal]{IEEEtran}
\usepackage{amsmath,amsfonts}
\usepackage{algorithmic}
\usepackage{algorithm}
\usepackage{array}
\usepackage[caption=false,font=normalsize,labelfont=sf,textfont=sf]{subfig}
\usepackage{textcomp}
\usepackage{stfloats}
\usepackage{url}
\usepackage{verbatim}
\usepackage{graphicx}
\usepackage{cite}
\usepackage{siunitx}
\usepackage{amsmath,amssymb,amsfonts}
\usepackage{algorithmic}
\usepackage{wrapfig}
\usepackage{booktabs}
\usepackage{threeparttable}

\begin{document}

\title{PhysDNet: Physics-Guided Decomposition Network of Side-Scan Sonar Imagery}

\author{Can Lei,
	Hayat Rajani,~\IEEEmembership{Member,~IEEE,},
	Nuno Gracias,
	Rafael Garcia,
	and Huigang Wang,~\IEEEmembership{Member,~IEEE}
\thanks{Can Lei and Huigang Wang are with the School of Marine Science and Technology, Northwestern Polytechnical University, Xi’an 710072, China. Huigang Wang is also with the Research \& Development Institute of Northwestern Polytechnical University in Shenzhen, Shenzhen 518057, China.}

\thanks{Hayat Rajani, Nuno Gracias and Rafael Garcia are with the Computer Vision and Robotics Research Institute (ViCOROB) of the University of Girona, Spain. This work was conducted while Can Lei was on a research stay at ViCOROB, Spain}

\thanks{This work was partly supported by the Spanish government through projects ASSiST (PID2023-149413OB-I00) and IURBI (CNS2023-144688). This work was also supported by the National Natural Science Foundation of China (62171368) and Science, Technology and Innovation of Shenzhen Municipality (JCYJ20241202124931042).}
\thanks{Corresponding author: Huigang Wang (e-mail: wanghg74@nwpu.edu.cn).}}



\maketitle

\begin{abstract}
Side-scan sonar (SSS) imagery is widely used for seafloor mapping and underwater remote sensing, yet the measured intensity is strongly influenced by seabed reflectivity, terrain elevation, and acoustic path loss. This entanglement makes the imagery highly view-dependent and reduces the robustness of downstream analysis. In this letter, we present PhysDNet, a physics-guided multi-branch network that decouples SSS images into three interpretable fields: seabed reflectivity, terrain elevation, and propagation loss. By embedding the Lambertian reflection model, PhysDNet reconstructs sonar intensity from these components, enabling self-supervised training without ground-truth annotations. Experiments show that the decomposed representations preserve stable geological structures, capture physically consistent illumination and attenuation, and produce reliable shadow maps. These findings demonstrate that physics-guided decomposition provides a stable and interpretable domain for SSS analysis, improving both physical consistency and downstream tasks such as registration and shadow interpretation.
\end{abstract}


\section{Introduction}

\IEEEPARstart{S}{ide}-scan sonar (SSS) has become an essential tool in underwater remote sensing, supporting applications such as seafloor mapping, ecological monitoring, and geological interpretation \cite{A2}. By providing large-scale acoustic backscatter imagery, SSS enables detailed characterization of seabed morphology and texture. At the same time, the recorded intensity is not a direct measure of seafloor properties but the outcome of multiple coupled factors including seabed reflectivity, terrain elevation, and acoustic path loss \cite{A3}. This dependency makes SSS imagery highly view-dependent and sensitive to acquisition geometry, which reduces physical interpretability and undermines the stability of downstream tasks such as registration, shadow analysis, and target recognition.

Most existing approaches to SSS analysis operate directly in the intensity domain. For registration and object classification, feature-based descriptors such as SIFT and AKAZE \cite{A4} and more recent deep neural networks \cite{A5} have been adapted to sonar imagery, but the strong view dependence of SSS causes the same object or region to appear very different across passes, making it difficult to extract stable and viewpoint-invariant features. Shadow segmentation faces similar challenges: geometry-based thresholding \cite{A6} and modern segmentation networks \cite{A7} have been employed, yet manual labeling is costly and multipath scattering in complex terrain often leads to ambiguous shadow regions that are hard to delineate. As a result, existing methods struggle to recover robust features for registration and classification, or to produce accurate shadow boundaries, highlighting the need to separate viewpoint-dependent effects from intrinsic seabed properties so that more stable features can be used in downstream tasks.

Analytical decomposition of SSS intensity into reflectivity, terrain, and propagation components has long been investigated \cite{A8}. However, the problem is highly non-linear and non-homogeneous, which makes closed-form solutions intractable and results extremely sensitive to noise and model assumptions. Traditional inversion methods are therefore difficult to apply in practice. With the rise of deep learning, several studies have attempted to approximate this inversion, but most approaches treat acoustic physics only as auxiliary constraints \cite{A9}, leaving the learned representations still entangled in the unstable intensity domain. A more promising direction is to embed imaging physics directly into the network itself, so that the learned decomposition remains both physically consistent and interpretable.

In this letter, we propose \textit{PhysDNet}, a physics-guided multi-branch network that decomposes SSS imagery into three interpretable components: seabed reflectivity, terrain elevation, and acoustic path loss. By embedding the Lambertian reflection model into the network, PhysDNet reconstructs sonar intensity from these disentangled fields, which enables self-supervised training without the need for ground-truth annotations. To further ensure physically plausible predictions, we introduce a progressive training strategy that incorporates weak priors from bottom-line measurements and shadow geometry. The main contributions are:  
\begin{itemize}
	\item A three-branch encoder–decoder architecture that disentangles raw sonar intensity into reflectivity, elevation, and path-loss fields, providing a stable and interpretable domain for analysis; 
	 
	\item A physics-consistent reconstruction process based on the Lambertian model, which enables self-supervised training without explicit labels;  
	
	\item A progressive training strategy that leverages weak physical priors from bottom-line measurements and shadow geometry, improving both physical consistency and generalization across environments.  
\end{itemize}

\section{Method}

\begin{figure*}[htbp]
	\centering
	\includegraphics[width=1\linewidth]{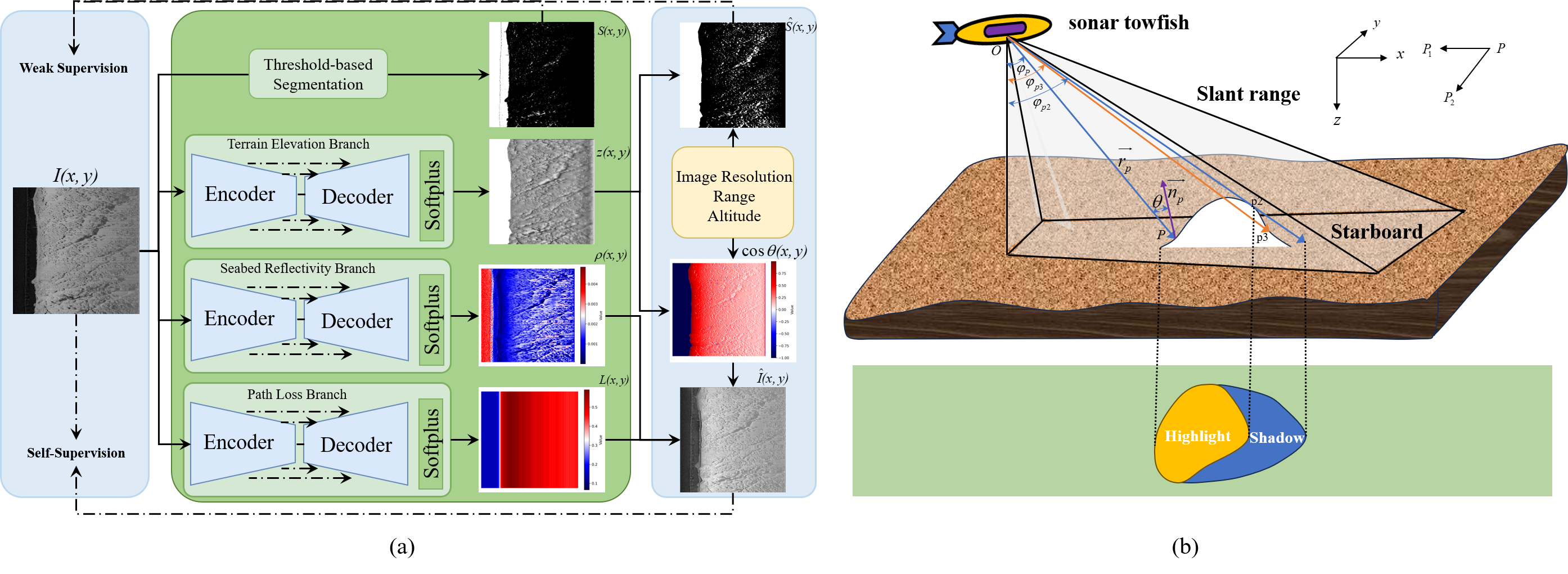}
	\caption{Overview of the proposed PhysDNet framework and the geometric definitions used in the physics-aware model. (a) PhysDNet employs a three-branch architecture to decouple SSS images into reflectivity ($\rho$), terrain elevation ($z$), and path loss ($L$), guided by the Lambertian model. A threshold-based shadow map $S(x,y)$ provides weak supervision, while predicted elevation supports physical computation of $\cos\theta$ and a physics-driven shadow map $\hat{S}(x,y)$. (b) At ping index $i$, the transducer at $O=(0, y_i, 0)$ emits toward seafloor points $P(x_{ij}, y_i, z_{ij})$. The reflection angle $\theta$ is defined between the acoustic incidence vector $\overrightarrow{OP}$ and the local surface normal (from neighbors $P_1$, $P_2$). The propagation angle $\varphi$ measures the grazing angle relative to the vertical axis. Points $P$, $p_2$, and $p_3$ illustrate the angular monotonicity rule for shadow detection, where $p_3$ lies in the acoustic shadow.}
	\label{fig01}
\end{figure*}

SSS intensity is strongly view-dependent due to anisotropic backscattering and shadowing, whereas seafloor structures (edges, contours, major geological formations) remain relatively stable across passes. Inspired by a Lambertian-like factorization for seabed imaging \cite{A8}, we decompose SSS into three physically interpretable fields (Fig. \ref{fig01}(a)): (i) seabed reflectivity $\rho$ (viewpoint-stable substrate property), (ii) the incidence factor $\cos\theta$ parameterized by terrain elevation $z$, and (iii) the propagation path-loss $L$. We propose \textit{PhysDNet}—a multi-branch encoder–decoder that jointly predicts $\rho$, $z$, and $L$ from a single SSS image—thereby enabling this decomposition.

\subsection{Multi-branch Physical Decoupling Network Architecture}

PhysDNet comprises three parallel branches with identical topology and independent parameters. Each branch takes the SSS intensity as input and outputs one physical map—$\rho$, $z$, or $L$—enabling explicit disentanglement. Because $L$ varies predominantly with range, we enforce range-wise consistency by averaging the predicted $L$ along columns and vertically replicating the result to a 2D field.

\begin{figure}[htbp]
	\centering
	\includegraphics[width=1\linewidth]{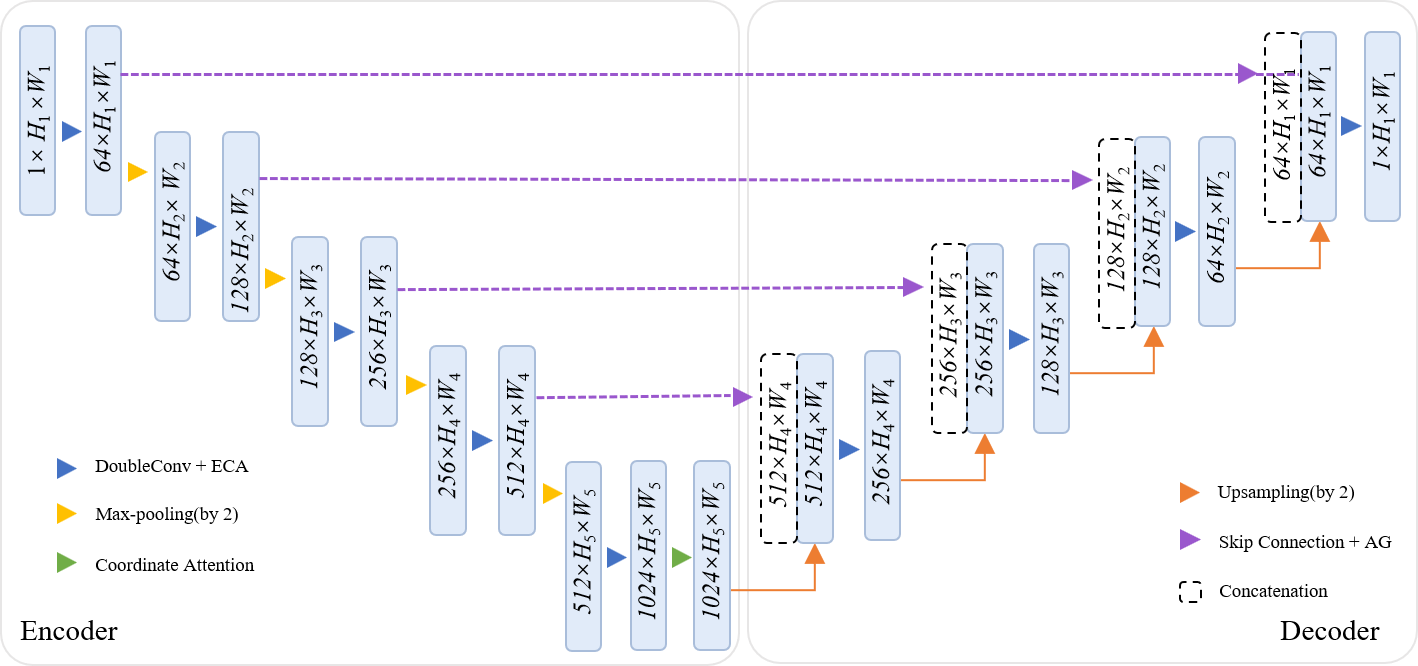}
	\caption{Detailed architecture of a single branch in PhysDNet. It consists of a DoubleConv for feature extraction, four downsampling stages (DoubleConv + MaxPooling), and four upsampling stages (Upsampling + Skip Connection + DoubleConv). Two AttentionBlocks are inserted at intermediate layers to enhance contextual features. The final OutConv produces the predicted physical map. Feature dimensions and module configurations are annotated in the diagram.}
	\label{fig02}
\end{figure}	

Each branch follows an enhanced Encoder-Decoder architecture: an initial \texttt{DoubleConv}, four encoder stages (\texttt{MaxPool + DoubleConv}), two \texttt{AttentionBlocks} at 256 and 512 channels, and a symmetric decoder with four upsampling stages using skip connections and \texttt{DoubleConv}. A final \texttt{OutConv} produces the target map; \texttt{Softplus} is applied to enforce non-negativity. This design preserves Encoder-Decoder architecture’s hierarchical capacity while adding branch-wise physical heads and attention to emphasize physically critical regions, laying the foundation for the physics-guided self-supervised learning introduced later (Fig.~\ref{fig02}).

\subsection{Reflection Angle Calculation and Intensity Reconstruction}

To reconstruct the sonar intensity image $\hat{I}(x,y)$ from the network outputs—terrain elevation $z(x,y)$, seabed reflectivity $\rho(x,y)$, and propagation loss $L(x,y)$—it is necessary to incorporate the reflection angle $\theta(x,y)$. According to the Lambertian reflection model \cite{A8}, the acoustic intensity is proportional to the cosine of the angle between the incident direction and the local seabed normal, formulated as:

\begin{equation}
	\hat{I}(x,y)=\rho (x,y)\cdot cos\theta (x,y)\cdot L(x,y)
\end{equation}

Accurately computing $\cos\theta(x,y)$ for each pixel is thus essential for reconstructing sonar intensity and supervising the network via physically grounded loss functions.

\subsubsection{Spatial Coordinate System Definition}

As illustrated in Fig. \ref{fig01}(b), we define a 3D spatial coordinate system to model acoustic propagation, where the x-axis aligns with the sonar range direction (i.e., image width), the y-axis follows the along-track direction (ping index), and the z-axis points downward. The origin $O = (0, 0, 0)$ is set at the transducer position at the start of scanning. At ping $i$, the transducer is located at $(0, y_i, 0)$, where $y_i = i \cdot \Delta y$ and $\Delta y$ is the fixed ping interval.
The $j$-th pixel at ping $i$ corresponds to a seafloor point $P = (x_{ij}, y_i, z_{ij})$, where $z_{ij}$ is the predicted seabed elevation. Since the SSS image stores slant range $r_{ij}$ along the image $x$-axis, the lateral coordinate is recovered by: ${{x}_{ij}}=\sqrt{r_{ij}^{2}-z_{ij}^{2}}$ ensuring subsequent geometry is consistent in 3D.

\subsubsection{Cosine of the Reflection Angle}
The incidence vector from the source to $P$ is $\overrightarrow{r_p}=\overrightarrow{OP}=(x_{ij},0,z_{ij})$. The local surface normal $\overrightarrow{n_p}$ is estimated from a tangent plane spanned by the left and previous-ping neighbors $P_1=(x_{i,j-1},y_i,z_{i,j-1})$ and $P_2=(x_{ij},y_{i-1},z_{i-1,j})$. The tangent vectors are
\begin{equation}
	\begin{array}{l}
		{{{\vec{v}}}_{1}}=\overrightarrow{{{P}_{1}}P}=({{x}_{ij}}-{{x}_{i,j-1}},0,{{z}_{ij}}-{{z}_{i,j-1}}) \\ 
		{{{\vec{v}}}_{2}}=\overrightarrow{{{P}_{2}}P}=(0,{{y}_{i}}-{{y}_{i-1}},{{z}_{ij}}-{{z}_{i-1,j}}) \\ 
	\end{array}
\end{equation}
and $\overrightarrow{n_p}=\vec{v}_1\times\vec{v}_2$. Finally, the cosine of the reflection angle is calculated via the normalized dot product:
\begin{equation}
	\cos {{\theta }_{p}}=\frac{\overrightarrow{{{r}_{p}}}\cdot \overrightarrow{{{n}_{p}}}}{\|\overrightarrow{{{r}_{p}}}\|\cdot \|\overrightarrow{{{n}_{p}}}\|}
\end{equation}

\subsubsection{Intensity Reconstruction}
The reconstructed map $\hat{I}(x,y)$ provides intermediate supervision by direct comparison to the raw SSS $I(x,y)$, guiding the network to learn seafloor geometry (via $\cos\theta$) and acoustic backscatter (via $\rho$) under physically consistent propagation ($L$).

\subsection{Shadow Map Generation}

Acoustic shadows—regions occluded from the sonar line of sight—encode strong constraints on seafloor relief. We derive a shadow map from the predicted elevation $z$ and use it as weak supervision to regularize $z$ in the absence of ground-truth heights.

\subsubsection{Propagation Angle}

Given the predicted elevation map $z(x,y)$ (with $z > 0$ representing seafloor depth), the 3D coordinate of a pixel at row $i$ (ping index) and column $j$ (range index) is defined as $P = (x_{ij}, y_i, z_{ij})$. We define the propagation angle $\varphi_{ij}$ between the acoustic ray from the transducer to the point $P$ and the vertical z-axis:
\begin{equation}
	\tan {{\varphi }_{ij}}=\frac{{{x}_{ij}}}{{{z}_{ij}}},\quad {{\varphi }_{ij}}=\arctan \left( \frac{{{x}_{ij}}}{{{z}_{ij}}} \right)
\end{equation}

\subsubsection{Shadow Detection via Angle Monotonicity}

Along each ping (fixed $i$), visibility must follow the nondecreasing envelope of $\varphi_{ij}$; terrain that raises the envelope occludes subsequent points with smaller grazing angles. We therefore classify $(i,j)$ as shadowed if:
\begin{equation}
	{{\varphi }_{ij}}<{{\max }_{k<j}}{{\varphi }_{ik}}
\end{equation}

This one-pass near-to-far scan per ping efficiently yields the shadow mask $\hat{S}(x,y)$ from $z(x,y)$. To illustrate this, Fig. \ref{fig01}(b) shows three sequential points $P$, $p_2$, and $p_3$ with $\varphi_P < \varphi_{p3} < \varphi_{p2}$. Here, $p_2$ is the last visible point before occlusion, while $p_3$, falling behind a steeper slope, is correctly identified as being in shadow.

\subsection{Loss Function Design and Training Strategy}

The proposed physically decoupled network adopts a multi-branch, multi-task architecture to jointly estimate the seafloor elevation map $z(x,y)$, bottom reflectivity $\rho(x,y)$, and acoustic path loss $L(x,y)$. In the absence of ground truth annotations for these physical quantities, we develop a three-stage training strategy, guided by sonar imaging physics and weak supervision signals.

\subsubsection{Stage 1: Initial Terrain Fitting and Row-wise Smoothness}

This stage aims to learn the coarse structure of the seafloor elevation. A weak reference elevation map $z_{\text{ref}}(x, y)$ is constructed by propagating the estimated bottom height from each ping, derived directly from the raw sensor data (e.g., altitude measurements).  This reference serves as an initial terrain prior to guide the prediction of $z(x,y)$. The training loss combines a elevation fitting term ${{\mathcal{L}}_{z\_fit}}$ and a smoothness regularization ${{\mathcal{L}}_{z\_row}}$:

\begin{equation}
	\begin{array}{l}
		{{\mathcal{L}}_{stage1}}={{\mathcal{L}}_{z\_fit}}+{{\mathcal{L}}_{z\_row}} \\ 
		{{\mathcal{L}}_{z\_fit}}=\|z(x,y)-{{z}_{ref}}(x,y){{\|}_{1}} \\ 
		{{\mathcal{L}}_{z\_row}}=\sum\limits_{y}{\sum\limits_{x}{\left| z(x+1,y)-z(x,y) \right|}} \\ 
	\end{array}
\end{equation}

The term $\mathcal{L}_{z\_row}$ encourages local continuity of the estimated elevation along the range ($x$) direction, consistent with the physical continuity of seafloor topography and the scanning geometry of side-scan sonar.

\subsubsection{Stage 2: Terrain Refinement with Shadow-guided Structural Consistency}

Building upon the coarse terrain estimated in Stage 1, this stage introduces vertical consistency, acoustic image fidelity, and weak supervision from shadow patterns. The overall loss integrates column-wise smoothness ${{\mathcal{L}}_{z\_col}}$, SSIM-based reconstruction ${{\mathcal{L}}_{ssim}}$, and shadow guidance ${{\mathcal{L}}_{shd}}$:

\begin{equation}
	\begin{array}{l}
		{{\mathcal{L}}_{stage2}}={{\mathcal{L}}_{z\_col}}+{{\mathcal{L}}_{ssim}}+{{\mathcal{L}}_{shd}} \\ 
		{{\mathcal{L}}_{z\_col}}=\sum\limits_{x}{\sum\limits_{y}{\left| z(x,y+1)-z(x,y) \right|}} \\ 
		{{\mathcal{L}}_{ssim}}=1-\text{SSIM}(I(x,y),\hat{I}(x,y)) \\ 
		{{\mathcal{L}}_{shd}}=\|S(x,y)-\tilde{S}(x,y){{\|}_{1}} \\ 
	\end{array}
\end{equation}

Analogous to $\mathcal{L}_{z\_\text{row}}$, $\mathcal{L}_{z\_\text{col}}$ enforces vertical (azimuth, $y$) smoothness of $z$, promoting cross-ping structural consistency. The SSIM term $\mathcal{L}_{\text{ssim}}$ drives the reconstruction $\hat{I}$ to preserve the structural characteristics of the original SSS image $I$. The shadow-trend loss $\mathcal{L}_{\text{shd}}$ provides weak supervision by comparing the predicted shadow map $\hat{S}$ with a coarse, thresholded mask $S$ from raw SSS; rather than enforcing pixel-level alignment, it encourages learning terrain structures that cause acoustic occlusions. Because shadows arise from elevation variations, $S$ indirectly refines $z$, and improved $z$ increases the reliability of the reflectivity and path-loss maps, which lack direct supervision.

\subsubsection{Stage 3: Joint Optimization and Semantic Enhancement}

This stage integrates all branches and promotes semantically meaningful reconstruction. The loss includes Elevation Smoothness Loss ${{\mathcal{L}}_{z\_smt}}$, Perceptual Loss ${{\mathcal{L}}_{perc}}$, Shadow Loss ${{\mathcal{L}}_{shd}}$, Path Loss Constraint ${{\mathcal{L}}_{pth}}$, Reflectivity Smoothness Loss ${{\mathcal{L}}_{\rho \_smt}}$, and combined loss ${{\mathcal{L}}_{stage3}}$:

\begin{small}
	\begin{equation}
		\begin{array}{l}
			{{\mathcal{L}}_{stage3}}={{\mathcal{L}}_{z\_smt}}+{{\mathcal{L}}_{perc}}+{{\mathcal{L}}_{shd}}+{{\mathcal{L}}_{pth}}+{{\mathcal{L}}_{\rho \_smt}} \\ 
			{{\mathcal{L}}_{z\_smt}}={{\mathcal{L}}_{z\_row}}+{{\mathcal{L}}_{z\_col}} \\ 
			{{\mathcal{L}}_{perc}}=\|\Phi (I)-\Phi (\hat{I})\|_{2}^{2} \\ 
			{{\mathcal{L}}_{pth}}=\|\text{mea}{{\text{n}}_{y}}(L(x,y))-\bar{I}(x){{\|}_{1}} \\ 
			{{\mathcal{L}}_{\rho \_smt}}=\sum\limits_{x,y}{\left| \rho (x+1,y)-\rho (x,y) \right|}+\left| \rho (x,y+1)-\rho (x,y) \right| \\ 
		\end{array}
	\end{equation}
\end{small}here, $\Phi$ extracts features from the first 10 layers of a pretrained MobileNetV3, so $\mathcal{L}_{\text{perc}}$ encourages learning deep semantic patterns. $\bar{I}(x)$ is the column-wise mean intensity of $I$, used to fit the predicted path-loss $L(x,y)$; thus $\mathcal{L}_{\text{pth}}$ enforces consistency with actual (range-dependent) acoustic attenuation. $\mathcal{L}_{\rho\_\text{smt}}$ regularizes spatial smoothness of $\rho$ to avoid spurious jumps, and $\mathcal{L}_{z\_\text{smt}}$ preserves terrain continuity. This staged training progressively learns terrain, enforces physical constraints, and injects semantic guidance; even without ground truth, the model leverages image structure, physical priors, and weak supervision to yield disentangled predictions of $z$, $\rho$, and $L$.

\section{Experiment}

\subsection{Data Description}

\subsubsection{Data Acquisition and Preprocessing}

We use SSS data from sector N07 of \cite{A10}, acquired with a Klein 3000H at \SI{500}{\kilo\hertz}; ranges were adjusted between \SI{50}{\meter} and \SI{100}{\meter} by bathymetry, with altitude \SI{10}{\percent} of range, yielding an average slant-range resolution of \SI{8.5}{\centi\meter}. The raw .xtf files provide intensities and metadata (altitude, range, USBL, heading). Preprocessing includes log-intensity normalization, port/starboard separation with port flipped, stacking pings into 2D waterfalls, and tiling into 1024×1024 patches with synchronized metadata for geometric modeling. The dataset is divided into 6734 training, 1704 validation, and 122 test samples, with the test set from regions distinct from the training and validation data.

\subsection{Experimental settings}

Experiments were run on Ubuntu with Python/PyTorch on a workstation with dual Intel\textsuperscript{\textregistered} Xeon\textsuperscript{\textregistered} Gold 6230 CPUs, 376\,GB RAM, and two NVIDIA Quadro RTX 6000 GPUs (CUDA 12.0, driver 525.147.05). Models were trained for 200 epochs (batch size 16) using Adam (lr=$1\times10^{-3}$, $\beta=(0.9,0.999)$, weight decay=$5\times10^{-4}$) with a cosine-annealing schedule and 30-epoch warmup, decaying the learning rate to $1\times10^{-6}$. Training follows a three-stage curriculum (epochs 0–60, 60–120, 120–200). Mixed precision, cuDNN acceleration, and gradient clipping are used for efficiency and stability.

\subsection{Experimental Results}

\subsubsection{Physical Decomposition Results} 

\begin{figure}[!t]
	\centering	\includegraphics[width=1\linewidth]{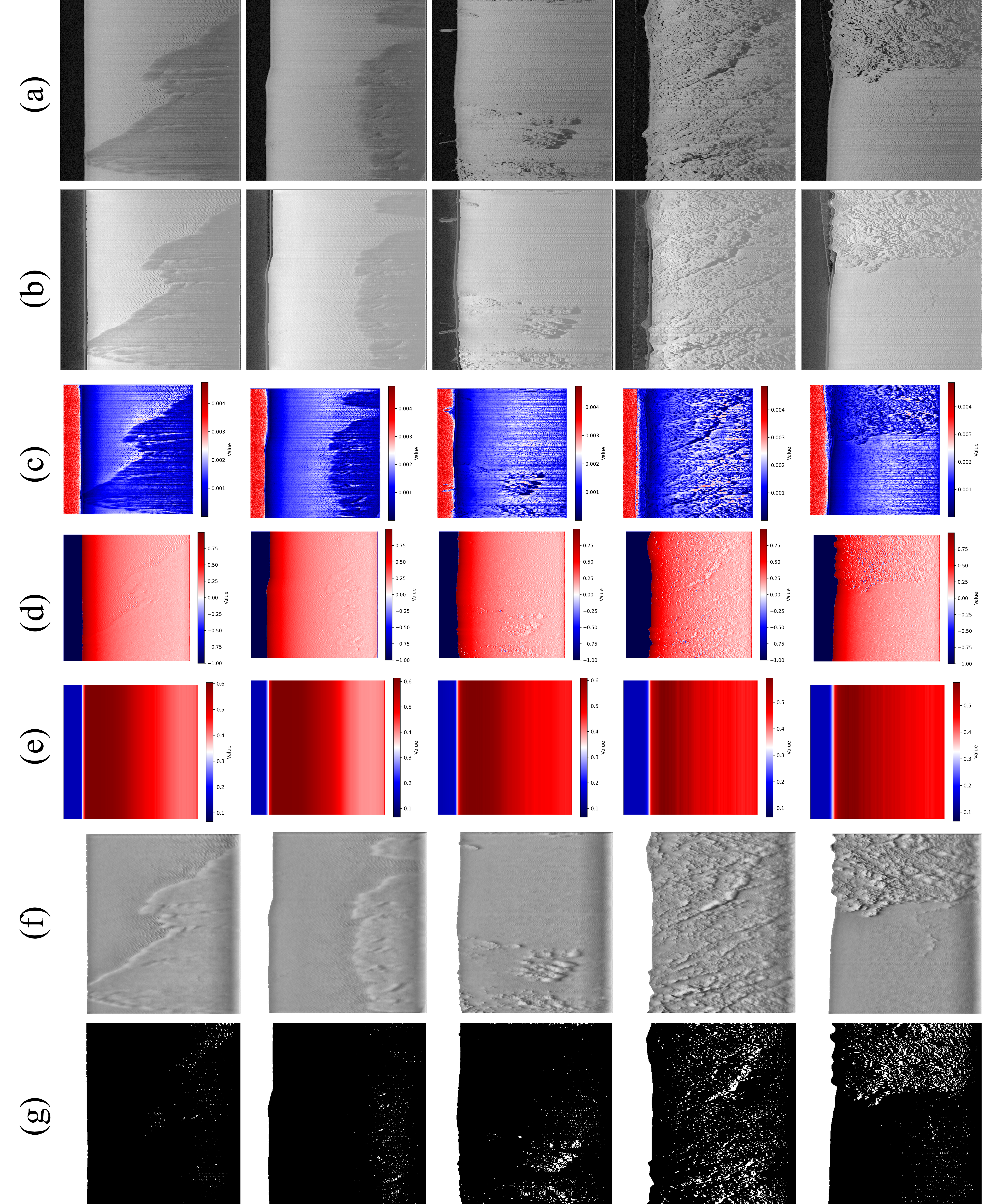}
	\caption{Visualization of the outputs from the multi-branch physically decoupled network. Each column corresponds to a test case, with 5 cases in total. From top to bottom: (a) raw side-scan sonar image; (b) predicted acoustic intensity map; (c) predicted seabed reflectivity map; (d) predicted cosine of the reflection angle ($\cos \theta $); (e) predicted acoustic path loss map; (f) predicted seabed elevation map (converted from depth); (g) shadow mask inferred from the predicted elevation.}
	\label{fig1}
\end{figure}

\begin{table}
	\caption{Quantitative results of our method using metric variables.}
	\vspace{1mm}
	\small
	\textbf{Note:} $S$—structural similarity between $\hat{I}(x,y)$ and $I(x,y)$; 
	$M$—mean squared error; $P$—perceptual distance. 
	$\bar e_{z}$—mean absolute error of elevation; $\sigma_{z}$—standard deviation of the error. 
	$\bar e_{L}$—mean absolute error of path-loss; $\sigma_{L}$—standard deviation of the error. All metrics are computed on the test set.
	
	\label{tab1}
	\centering
	\footnotesize
	\setlength{\tabcolsep}{10.5pt}
	\renewcommand{\arraystretch}{1.2}
	\begin{tabular}{ccccccc}
		\toprule
		 $S$ & $M$ & $P$ & $\bar e_{z}$ & $\sigma_{z}$ &  $\bar e_{L}$ & $\sigma_{L}$ \\
		\midrule
		 0.83 & 0.04 & 0.02 & 0.32 & 0.07 & 0.007 & 0.004 \\
		\bottomrule
	\end{tabular}
\end{table}

As shown in Fig. \ref{fig1} and Table \ref{tab1}, the reconstructed intensity maps closely approximate the original SSS, reproducing the spatial backscatter distribution ($S{=}0.83$, $M{=}0.04$, $P{=}0.02$). The reflectivity maps suppress view-dependent fluctuations while preserving stable geological structures, confirming the viewpoint-invariant property of $\rho$. The cosine maps capture illumination variation, and the path-loss maps show the expected monotonic attenuation with range, with strong spatial coherence and low error ($\bar e_{L}{=}0.007$, $\sigma_{L}{=}0.004$). The elevation maps follow seafloor undulations and interpolate smoothly across shadows, yielding bounded error ($\bar e_{z}{=}0.32$ m, $\sigma_{z}{=}0.07$ m). In addition, terrain underestimation derived from $z$ delineates shadow onsets and transitions, producing masks with cleaner boundaries and fewer omissions than heuristic thresholding.

\subsubsection{Shadow-Based Evaluation} 

\begin{figure}[!t]
	\centering	\includegraphics[width=1\linewidth]{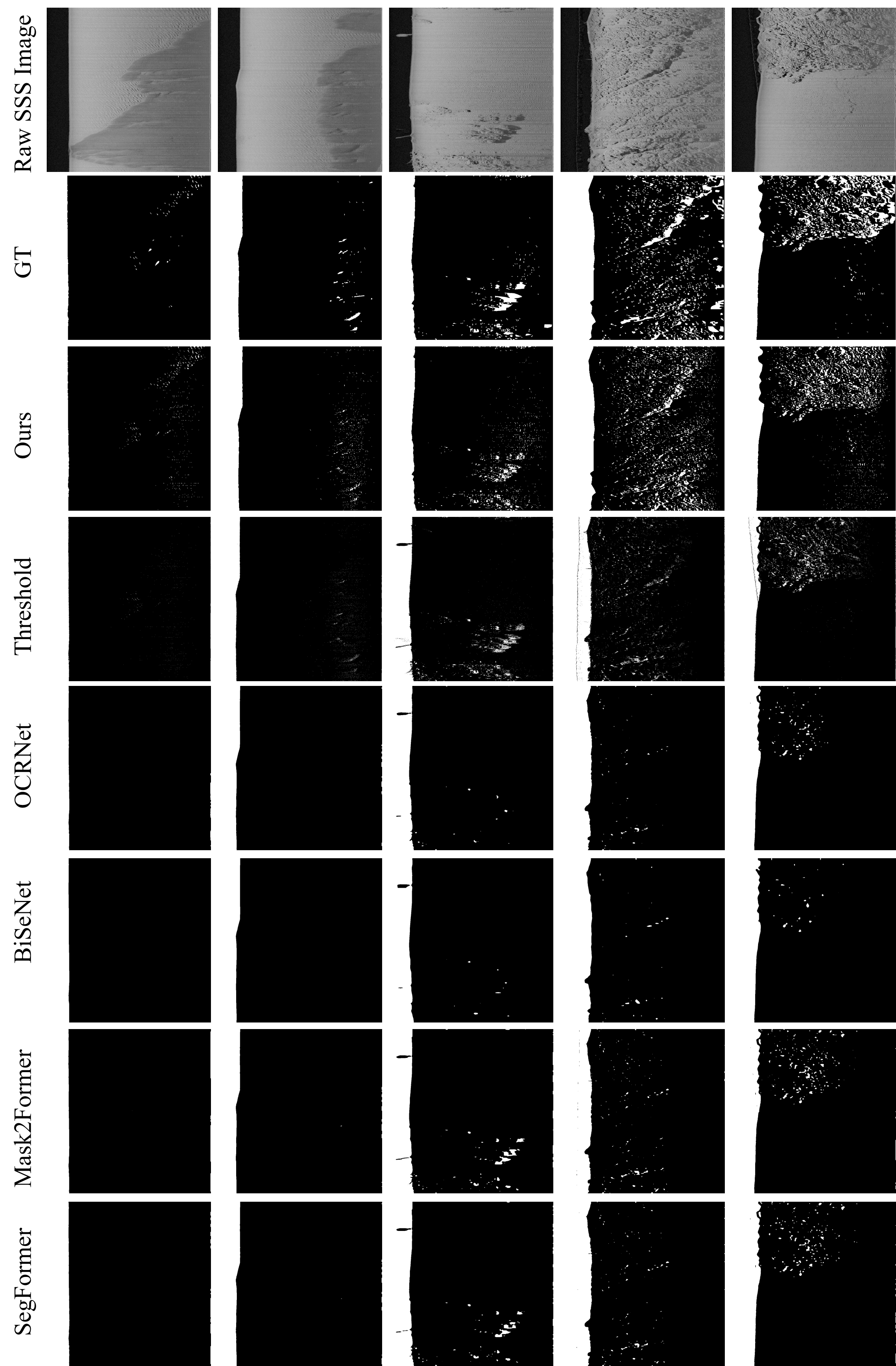}
	\caption{Shadow segmentation comparison. From top to bottom, input SSS, manual GT, ours (shadows derived from predicted $z$), adaptive threshold, OCRNet, BiSeNet, Mask2Former, SegFormer. Our method best matches GT and recovers fine low contrast shadows.}
	
	\label{fig2}
\end{figure}	

\begin{table}
	\caption{Shadow segmentation performance on the test set.}
	\vspace{1mm}
	\small
	\textbf{Note:} IoU denotes intersection over union; ACC denotes pixel accuracy; P denotes precision; R denotes recall.
	
	\label{tab2}
	\centering
	\footnotesize
	\setlength{\tabcolsep}{8.5pt}
	\renewcommand{\arraystretch}{1.2}
	\begin{tabular}{ccccc}
		\toprule
		\textbf{Method} & \textbf{IoU} & \textbf{ACC} & \textbf{P} & \textbf{R}  \\
		\midrule
		Threshold \cite{A6}  & 77.6\%   & 89.17\% & 87.99\%  & 78.74\%  \\
		OCRNet \cite{A12}  & 75.89\% & 87.99\% & 86.67\% & 76.02\% \\
		BiSeNet \cite{A13}  & 75.84\% & 87.92\% & 86.60\% & 75.86\% \\
		Mask2Former \cite{A14} & 77.15\% & 88.66\% & 87.37\% & 77.41\% \\
		SegFormer \cite{A15} & 77.06\%  & 88.61\% & 87.29\%  & 77.29\% \\
		\textbf{Ours} & \textbf{92.64\%}  & \textbf{96.97\% } & \textbf{96.40\%} & \textbf{94.49\%} \\
		\bottomrule
	\end{tabular}
\end{table}

Since no public ground truth exists for $z, \rho,$ and $L$, we validate the decomposition in the shadow domain, where shadows are derived from predicted $z$ through imaging geometry. We compare against adaptive thresholding \cite{A6} and five representative segmentation baselines, including OCRNet \cite{A12}, BiSeNet \cite{A13}, Mask2Former \cite{A14}, and SegFormer \cite{A15}. All deep baselines are trained on the same threshold-generated masks, and performance is evaluated on a manually annotated test set using IoU, ACC, Precision, and Recall.

As shown in Fig.~\ref{fig2} and Table~\ref{tab2}, our method best matches ground truth, recovering fine-scale and low-contrast shadows that other methods consistently miss. SegFormer and Mask2Former perform best among the baselines but still fail on small shadow structures, while others capture only coarse blobs. Quantitatively, our method achieves IoU 92.64\%, ACC 96.97\%, Precision 96.40\%, and Recall 94.49\%, with the most notable gain in recall ($+17.1\%$ over the best deep baseline), reflecting its superior recovery of small and subtle shadows.

\section{Conclusion}

In this letter, we introduced PhysDNet, a physics-guided multi-branch network that disentangles SSS imagery into seabed reflectivity, terrain elevation, and acoustic path loss. By embedding acoustic imaging physics into the network, the framework enables self-supervised training and provides interpretable representations that improve the robustness of downstream analysis. Experiments showed that PhysDNet produces consistent decompositions, with reflectivity maps preserving viewpoint-invariant structures, elevation maps offering reliable shadow cues, and path-loss maps following expected attenuation trends. However, the current framework is limited by its reliance on a Lambertian reflection approximation, which simplifies acoustic scattering and may reduce accuracy under complex seabed conditions. Future work will focus on incorporating more advanced scattering models to enhance physical fidelity while maintaining interpretability.

\bibliographystyle{IEEEtran}
\bibliography{refernew}

\end{document}